\documentclass[a4paper]{article}
\usepackage[left=4cm,right=3cm,top=3cm,bottom=3cm]{geometry}

\usepackage{amsmath,amssymb,graphicx,url}

\begin{document}
\author{Kevin F. S. Pardede$^{1,\dagger}$, Agus Suroso$^{1,2,*}$, Freddy P. Zen$^{1,2,\#}$\\[0.2cm]
$^1$Theoretical Physics Laboratory, THEPI Division,\\
Institut Teknologi Bandung, Jl. Ganesha 10 Bandung 40132, Indonesia\\
$^2$Indonesia Center for Theoretical and Mathematical Physics (ICTMP),\\ Bandung, Indonesia\\
\small{Email: $^\dagger$\url{kfspardede@gmail.com}, $^*$\url{agussuroso@fi.itb.ac.id}, $^\#$\url{fpzen@fi.itb.ac.id}}
}
\title{Nonlinear matter terms in general scalar-tensor braneworld cosmology}
\date{}

\maketitle

\begin{abstract}
A five dimensional braneworld cosmological model in general scalar-tensor action comprises of various Horndeski Lagrangian is considered. The Friedmann equations in the case of the strongly and weakly coupled $\mathcal{L}_5$ Horndeski Lagrangian have been obtained. The strongly coupled $\mathcal{L}_5$ model produces the Cardassian term $\rho^n$ with $n=\pm 1/2$, which can served as alternative explanation for the accelerated expansion phase of the universe. Furthermore, the latest combined observational facts from BAO, CMB, SNIa, $f_{\sigma_8}$, and $H_0$ value observation suggest that the $n=-1/2$ term lies quite close to the constrained value. On the other hand, the weakly coupled $\mathcal{L}_5$ case has several new correction terms which are omitted in the braneworld Einstein-Hilbert model, e.g. the cubic $\rho^3$ and the dark radiation-matter interaction term $ \chi a^{-4}\rho$. Furthermore, this model provides a cosmological constant constructed from the bulk scalar field, requires no brane tension, and supports the big bang nucleosynthesis (BBN) constraint naturally. \\[0.2cm]

\end{abstract}

\section{Introduction}
\label{intro}
General relativity (GR) remains among the most successful physics theory at the moment and still attracts interest of many physicists today. For example, in 2015, LIGO's finding on gravitational waves from binary black hole merger that occurred billions of years ago \cite{grav_waves_LIGO} proved the consistency of the theory.  

Despite of its enormous success, GR has some limitations. The most notable one is the inconsistency with the fact that the universe undergoes an accelerated expansion phase \cite{riess1998observational,perlmutter1999measurements}. Cosmological constant is the simplest modification which solve that inconsistency, but the theoretical prediction is still dramatically off from the observation i.e. by factor of $10^{120}$.

Another explanation for the late acceleration has been given by Freese and Lewis in 2002 \cite{freese2002cardassian}, where they considered the following modified Friedmann equation
\begin{equation}\label{cardassian_ansatz_friedmann}
H^2 = A \rho + B \rho^n,
\end{equation}
where $A \equiv \kappa_4^2/3$ and $B$ is some constant. In the case where $n<2/3$ the $\rho^n$ term is called the Cardassian term \cite{freese2002cardassian} and can be shown to provide an accelerated expansion phase in the matter dominated era. The Cardassian Friedmann equation \eqref{cardassian_ansatz_friedmann} can be generalized into
$H^2 = g(\rho)$, where $g(\rho)$ is some arbitrary function that is approximately  $\rho$ beyond some threshold high energy scale while still producing the accelerated expansion in the matter dominated area. One of the most common models is called the polytropic Cardassian \cite{wang2003future}
\begin{equation}\label{cardassian_ansatz_friedmann_poly}
H^2 = A \rho \left[1+ \left({\frac{\rho_{\text{car}}}{\rho}}\right)^{m(1-n)}\right]^{1/m},
\end{equation}
where $\rho_{\text{car}}$ satisfy $A\rho_\text{car} = B \rho_\text{car}^n$. It can be seen that when $m=1$, equation \eqref{cardassian_ansatz_friedmann_poly} went back to \eqref{cardassian_ansatz_friedmann}. 

The Cardassian term has been noted by Chung, et al. in Ref.  \cite{chung1999cosmological} to be originated from some specific bulk energy momentum tensor $T_{AB}$ in braneworld scenario.  Braneworld scenario is basically a specific scenario when we modify GR by considering a higher (larger than four) dimensional spacetime. This approach is motivated by some candidates of unification theory such as M-theory which predicts that our spacetime is an eleven dimensional manifold. In particular, Horava-Witten solution suggests that the spacetime is $\mathbb{R}^{10} \times S^1/\mathbb{Z}_2$, where six of the extra dimensions were compactified on a very small scale leading to an effectively five dimensional theory\cite{maartens_2010} $\mathbb{R}^{4} \times S^1 / \mathbb{Z}_2$ . This assumption leads to a model called the braneworld model where our 3+1 spacetime is a hypersurface called brane in a five dimensional bulk \cite{randall_1,randall_2}.

In particular, the first and second modified Friedmann equations in the braneworld model with Einstein-Hilbert action read \cite{binetruy_2}
\begin{align}
\left(\frac{\dot{a}}{a}\right)^2 &= \frac{\kappa_4^2}{3} \rho - \frac{k}{a^2} + \frac{\Lambda_4}{3} +  \frac{\kappa_5^4}{36} \rho^2  + \frac{\chi}{a^4}, \label{friedmann_1_braneworld_EH} \\
\frac{\ddot{a}}{a} &= - \frac{\kappa_4^2}{6} (3 p + \rho) + \frac{\Lambda_4}{3} - \rho^2 \frac{\kappa_5^4}{18}  - \rho p \frac{\kappa_5^4}{12}  - \frac{\chi}{a^4}, \label{friedmann_2_braneworld_EH}
\end{align}
where $\chi$ is a constant,  $\sigma$ is the brane tension, $\Lambda_4 \equiv \tfrac{1}{2}\kappa_5^2(\Lambda_5 + \tfrac{1}{6} \kappa_5^2 \sigma^2)$ is the effective four dimensional cosmological constant resulting from the bulk cosmological constant $\Lambda_5$, $\kappa_4^2 \equiv \tfrac{\sigma}{6}\kappa_5^4$ is the Einstein's four dimensional kappa constant, where $\kappa_5$ is the Einstein's five dimensional kappa constant.

The last two terms of \eqref{friedmann_1_braneworld_EH} are the correction terms for the Friedmann equation. As can be seen, the braneworld model gives rise to both high energy quadratic matter term, and low energy correction from the dark radiation term $\chi a^{-4}$ and effective cosmological constant $\Lambda_4$. It can also be inferred that the brane tension $\sigma$ contributes to the effective cosmological constant $\Lambda_4$ on the brane, so that cosmological constant on the bulk is not mandatory anymore to provide the acceleration phase. It also can be seen that a nonzero brane tension is necessary for the modified Friedmann equation to be consistent with its low energy limit, which is the conventional four dimensional Friedmann equation.

Another popular way to modify GR is to consider a scalar-tensor theory where the metric tensor and its derivatives are coupled to a scalar field $\phi$. In this context a healthy theory is such that the Lagrangian doesn't contain nondegenerate higher (than two) derivative Lagrangian, which can suffer from the Ostrogradsky instability. The most general scalar-tensor theory in four dimensional spacetime that produces second order field equations is called the Horndeski theory. The Horndeski theory is built up from four Lagrangian ($\mathcal{L}_i$,  $i=2,3,4,5$). By appropriately choosing the parameters in each Lagrangian, well known specific scalar-tensor Lagrangian can be recovered, such as quintessence \cite{caldwell1998cosmological}, k-essence \cite{armendariz2000dynamical}, nonminimal derivative coupling (NMDC) \cite{amendola1993cosmology,sushkov_2009,suroso2013,suroso2012,Hikmawan:2016tif}, etc. are all contained in Horndeski theory.   

The theory has been discovered by Horndeski a long time ago in 1974 \cite{horndeski1974second} but has gained a lot of attention only recently in light of its equivalence to covariantization of Galileon \cite{deffayet2009covariant,deffayet2009generalized,deffayet2011k}. Galileon is a scalar field invariant under the "Galilean" transformation $\phi \rightarrow \phi + a + b_{\mu} x^\mu$, where $a, b_{\mu}$ are some constants. The Galileon is interesting because it has been found that DGP braneworld model \cite{deffayet2002accelerated} which provides a self-accelerating cosmology, under certain decoupling limit can be described as a scalar-tensor Lagrangian which has second order field equations \cite{gabadadze2006coupling} and such that the scalar field possesses the aforementioned "Galilean" symmetry \cite{nicolis2009galileon}.  

As we have seen, braneworld model produces nonlinear matter term. Naturally, we can ask whether action different from Einstein-Hilbert in braneworld model can produce another different nonlinear matter terms, or even the sought after Cardassian term. In particular, in this article, cosmological model in a five dimensional braneworld model which action originates from a general scalar-tensor theory comprises of various Horndeski Lagrangian is investigated. There are several works along this line of research. Friedmann equations resulting from braneworld model with Einstein-Hilbert Lagrangian have been obtained by several authors a long time ago in 1999 (to name a few, see Refs. \cite{sms,binetruy_1,ida2000brane}). Dynamical analysis and the effect of the dark radiation term have been investigated in \cite{Pardede:2018nsg}. Cosmology in braneworld model with Lorentz invariant violation has been reported in \cite{zen2010modified}.  Kaluza-Klein brane cosmology has been considered in the case of one brane \cite{zen_kaluzakleinbrane} and two brane \cite{feranie_4plusn}. Braneworld with minimal coupling bimetric cosmology has been considered in Ref. \cite{bimetric_brane_youm}, while the nonminimal coupling braneworld has been considered in Ref. \cite{nonminimal_scalar_bogdanos}.  A nonminimal derivative coupling (NMDC) five dimensional braneworld model has been investigated by several authors, e.g. in the case of time dependent scalar field \cite{suroso2013,widiyani2015randall} or where the scalar field is the function of extra dimension coordinate \cite{minamitsuji_2014}.

In the next section, the field equations for cosmology in the general scalar-tensor braneworld theory will be derived. After that, we derive and analyze the consequences of the modified Friedmann equations for the cases where the fifth Horndeski Lagrangian $\mathcal{L}_5$ is either strongly or weakly coupled to the rest of the Lagrangian. For the strongly coupled case we have identified Cardassian terms $\rho^n$ for $n=\pm1/2$.  In the weakly coupled case we found higher order matter cubic term and the dark radiation-matter interaction term. Finally, as in Ref. \cite{Pardede:2018nsg} for the case of braneworld model, we will investigate how those correction terms affect the evolution of the universe through the Hubble diagram.

Before we continue, we would like to give a quick remark about the observational constraint on this theory.  For example, gravitational wave observation  GW170817 \cite{GW170817} has confirmed that gravitational waves travel with the speed of light with deviation smaller than $10^{-15}$. By using the language of effective field theory of dark energy, this  measurement implied some precise relations among the operators \cite{creminelli_dark}. In particular, for example it can be shown from the relations that $\mathcal{L}_5$ is excluded.  However, it should be noted that the energy scale of this event lie very close to typical cutoff of many DE models. Therefore, the validity of this constraint is still a subject of debate, because the UV completion can modify the speed of the gravitational waves \cite{rainbow}. Moreover, if we consider a possibility that the speed of gravitational waves can vary in time, it can dynamically set to unity at present without introducing any fine tuning between the operators in Horndeski\cite{copeland2019dark}. By considering the previous "loophole" it  can be argued that even nontrivial contribution from $\mathcal{L}_5$ can be compatible with the data \cite{copeland2019dark}. In this regard, we will assume that $\mathcal{L}_5$ is not ruled out convincingly by the gravitational waves experiment yet.

\section{Setup and field equations}
The setup is a braneworld model with the following metric
\begin{equation}\label{metric}
ds^2 = g_{AB} dX^A dX^B = q_{\mu \nu} dx^\mu dx^\nu + dy^2, 
\end{equation}
where $y$ is the extra dimension coordinate (the brane is located at $y=0$) and $g_{AB}$, $q_{\mu \nu}$ is the five and four dimensional metric respectively. The four dimensional metric reads
\begin{equation}
q_{\mu \nu} dx^\mu dx^\nu = -N^2(t,y) dt^2 + a^2(t,y) \gamma_{ij} dx^i dx^j,
\end{equation}
where $N$ is some function, $a$ is scale factor, and $\gamma_{ij}$ is the three dimensional maximally symmetric space for which $k=-1,0,1$ refers to hyperbolic, flat and spherical space respectively. Some words about notation, in this article braneworld model is $4+1$ dimensional,   where the full spacetime coordinate is denoted as $X^A = (X^0, X^1, \dots , y \equiv X^4)$, brane coordinate as $x^\mu = (x^0, x^1, \dots x^3)$, and spatial coordinate as $x^i = (x^1, x^2, x^3)$. We denote $\dot{f} = df/dt$ and $f' = df/dy$ for some function $f$. Lastly, we will occasionally refer to $R$ and ${}^{(q)}R$ for example, as five and four dimensional Ricci scalar, and similarly for any other tensor.

The action in this model is 
\begin{equation}
S= \int \frac{d^5 X}{2 \kappa_5^2} \sqrt{-g}(R + \mathcal{L}_H) + S_{b},
\end{equation}
where $\kappa_5^2 \equiv 8\pi G_5$ is the five dimensional Einstein's kappa constant, where the gravitational constant is denoted by $G_5$. $\mathcal{L}_H$ is the five dimensional Horndeski Lagrangian
\begin{equation}\label{mathcal_L_H}
\mathcal{L}_H \equiv \sum_{i=2}^5 \xi_i \mathcal{L}_i,
\end{equation}
where $\xi_i$'s are the coupling constants while the $\mathcal{L}_i$'s are defined as follows
\begin{equation}\label{Horndeski_Lagrangian}
\begin{aligned}
\sum_{i=2}^{5} \mathcal{L}_i  &= G_2(\phi, X)  \\
&~~ + G_3 (\phi, X)  \square \phi \\
&~~+ G_4(\phi, X)  R + G_{4X} \left[(\square \phi)^2 - \phi_{AB} \phi^{AB}\right] \\
&~~+ G_5(\phi, X)  G_{A B} \phi^{AB} - \frac{1}{6} G_{5X}\left[(\square \phi)^3 - 3 (\square \phi ) \phi_{A B} \phi^{AB} + 2 \phi_{AB} \phi^{BC} {\phi_{C}}^A\right] ,
\end{aligned}
\end{equation}
with an obvious classification for $\mathcal{L}_i$, where $\phi_A \equiv \nabla_A \phi$, $X \equiv \nabla_A \phi \nabla^A \phi$, and $G_{iX} = \partial G_i/\partial X$, $G_{i\phi} = \partial G_i/\partial \phi$ for $i=2,3,4,5$.
In the definition above, $S_b$ is the matter in the brane action
\begin{equation}
S_b = \int d^4x \sqrt{-q} \mathcal{L}_b[q,\phi],
\end{equation}
which variation with respect to four dimensional metric, comprises of brane tension $\sigma$ and brane energy momentum tensor $\tau_{\mu \nu}$
\begin{equation}\label{S_mu_nu}
S_{\mu \nu} \equiv - \frac{2}{\sqrt{-q}} \frac{\delta }{\delta q^{\mu \nu}} (\sqrt{-q} \mathcal{L}_b) = - \sigma q_{\mu \nu} + \tau_{\mu \nu}.
\end{equation}
Perfect fluid is chosen as the brane energy momentum tensor.

In this section we will use variational method to obtain the field equations. In order to do so, first we need to transform the Lagrangian \eqref{Horndeski_Lagrangian} into geometric form, for which $R$, $G_{AB}$, and $\phi_{AB}$ are translated into brane variables such as the brane Ricci scalar ${}^{(q)}R$ and extrinsic curvature $K_{\mu \nu} =  {q_{\mu}}^A {q_\nu}^B \nabla_A n_B$, where $n^A$ is a normal vector with respect to the brane. 

For the Einstein-Hilbert action, the result is the well known projection identity of Ricci scalar \cite{maeda_einstein_brane}
\begin{equation}\label{Ricci_scalar_projection}
R = {}^{(q)} R + K^2 - K^{AB}K_{AB} + 2 \nabla_A (n^B \nabla_B n^A - n^A \nabla_B n^B),
\end{equation}
where in the braneworld case, the normal vector is spacelike $n_A n^A = 1$.
For the Horndeski Lagrangian, the translation procedure has been carried out by Gleyzes, et al. \cite{gleyzes2013essential} in the case of four dimensional spacetime, where the "brane" is taken to be the $\phi$ constant hypersurface. In Ref. \cite{gleyzes2013essential}, the scalar field is assumed to be a function of $t$ only, so the hypersurface considered in that paper is a time constant hypersurface. Interestingly enough, this procedure also works in our case where our brane is a $y$ constant hypersurface, if we assume that our scalar field is a $y$ only dependent function, $\phi = \phi(y)$. Following the aforementioned procedure, assuming $\phi = \phi(y)$, the Horndeski Lagrangian \eqref{Horndeski_Lagrangian} can be translated into a geometric form as follows
\begin{equation}\label{Horndeski_Lagrangian_geometric}
\begin{aligned}
\sum_{i=2}^{5} \mathcal{L}_i &= G_2(\phi, X)  \\
&~~-2X^{3/2} K F_{3X} - F_{3\phi} X \\
&~~+ G_4{}^{(q)}R - (K^2 - K_{AB} K^{AB})(2G_{4X} X - G_4) + 2G_{4\phi} K X^{1/2} \\
&~~+ \Bigg[F_5 X^{1/2} \left(K^{AB} {}^{(q)}R_{AB} - \frac{K}{2} {}^{(q)}R \right) + \frac{X}{2}(G_{5\phi}-F_{5\phi}) ~{}^{(q)}R
\\&~~+ \frac{X}{2}  G_{5\phi} (K^{AB} K_{AB} - K^2) \\&~~+ \frac{G_{5X}}{3} X^{3/2} \left(K^3 - 3 K K_{AB} K^{AB} + 2K_{AB} K^{AC} {K_C}^B\right)\Bigg],
\end{aligned}
\end{equation}
where $F_3$ and $F_5$ are defined as follows
\begin{equation}
\begin{aligned}
& G_3 \equiv F_3 + 2X F_{3X},\\
& G_{5X} \equiv F_{5X} + \frac{F_5}{2X},
\end{aligned}
\end{equation}
and $F_{iX} = \partial F_i/ \partial X$, $F_{i\phi} = \partial F_i/\partial \phi$.

In the rest of this article, we derive the modified Friedmann equations for the Horndeski Lagrangian \eqref{Horndeski_Lagrangian_geometric}, albeit with some assumptions. Firstly, for the reason that will be obvious when we calculate the scalar field equation, let us assume that $G_{i \phi} = 0$, for $i=2,3,4,5$.  Next, as we will show, unlike the braneworld model with Einstein-Hilbert action \cite{sms}, the Friedmann equation resulting from a model that contains the $\mathcal{L}_3$ will be able to go back to the conventional four dimensional Friedmann equation, even with zero brane tension. Thus we set $\sigma=0$ in our definition of brane matter action \eqref{S_mu_nu}. In short, we consider the following action
\begin{equation}\label{action_after_assumption}
\begin{aligned}
S &=\int \frac{d^5 X}{2 \kappa_5^2} \sqrt{-g} \left( R + \xi_2 \mathcal{L}_2 + \xi_3 \mathcal{L}_3 + \xi_4 \mathcal{L}_4 + \xi_5 \mathcal{L}_5\right) + S_{b} \\
&= \int \frac{d^5 X}{2 \kappa_5^2} \sqrt{-g} \Bigg\{R + \xi_2 G_2 - 2\xi_3 \phi'^3 K F_{3X}  \\&~~+ \xi_4 \Big(G_4 {}^{(q)}R - (K^2 - K_{\mu \nu} K^{\mu \nu})(2G_{4X} \phi'^2 - G_4)\Big) \\&~~+ \xi_5\Bigg[F_5 \phi' \left(K^{\mu \nu} {}^{(q)}R_{\mu \nu} - \frac{K}{2} {}^{(q)}R \right) \\&~~+ \frac{G_{5X}}{3} \phi'^3 \left(K^3 - 3 K K_{\mu \nu} K^{\mu \nu} + 2K_{\alpha \beta} K^{\alpha \gamma} {K_\gamma}^\beta\right)\Bigg] \Bigg\} + S_b.
\end{aligned}
\end{equation}

\section{Cardassian terms from strong coupling $\mathcal{L}_5$ Friedmann equation}

In this section we compute bulk field equations and the junction conditions for general action \eqref{action_after_assumption}. After that, we apply the strong $\mathcal{L}_5$ coupling condition to find the modified Friedmann equations.

Let us first compute the scalar field equation. From the variation of \eqref{action_after_assumption} with respect to $\phi$ we have
\begin{equation}\label{scalar_field_eq}
\begin{aligned}
\mathcal{C}(t) &= -2\xi_2a^3N G_{2X} \phi' + 2\xi_3 \phi'^2\left(3F_{3X} + 2\phi'^2F_{3XX}\right)(3a'a^2 N + a^3N')\\&-12\xi_4 \Bigg[ G_{4X} \phi' \left(kaN + \frac{\ddot{a} a^2}{N} + \frac{\dot{a}^2 a}{N} - \frac{\dot{N} \dot{a} a^2}{N^2}\right) \\
&~~~~~~-\phi'\left(2G_{4XX}\phi'^2+G_{4X} \right)(a^2 a'N' + aa'^2N)\Bigg] \\&+\xi_5\Bigg[3\left(2F_{5X}\phi'^2 + F_5 \right)\left(\frac{a' \dot{a}^2}{N} - 2 \frac{\ddot{a} a' a}{N}+ ka'N + k a N' + \frac{\dot{a}^2 a N'}{N^2} -2 \frac{a' \dot{a} a \dot{N}}{N^2}\right) \\
&~~~~~+2\phi'^2\left(2G_{5XX}\phi'^2 + 3G_{5X}\right)(a'^3 N + 3a'^2 a N')\Bigg],
\end{aligned}
\end{equation}
where $\mathcal{C}(t)$ is some function. Notice now that our $G_{i\phi} = 0$ assumption has ensured that the scalar field equation is a first order differential equation in $\phi$.

Next, we will derive the $yy$ and $\mu y$-field equations. To do that, we need to introduce shift scalar $b$ and shift vector $b^\mu$ into our metric as follows \cite{gao2010modified}
\begin{equation}\label{metric_adm}
ds^2 = b^2 dy^2 + q_{\mu \nu} (dx^\mu + b^\mu dy)(dx^\nu + b^\nu dy).
\end{equation}
Note that in this metric 
\begin{align}
X &= \frac{\phi'^2}{b^2}, \\
K_{\mu \nu} &= \frac{1}{2b} \left(\partial_y q_{\mu \nu} - {}^{(q)} \nabla_\mu b_\nu - {}^{(q)} \nabla_\nu b_\mu \right) \label{curv_ex_adm}.
\end{align}
After we have the field equations on our hand, we can set $b=1$, $b^\mu = 0$ to obtain our original metric \eqref{metric}. By varying the action \eqref{action_after_assumption} with respect to $b^\mu$, we obtain the $\mu y$-field equation. From \eqref{curv_ex_adm}, we can see that $b^\mu$ only appear in terms containing $K_{\mu \nu}$. After some algebra, it can be shown that by assuming $b$ constant, the $ty$-equation ($\mu=0$) can be solved by taking
\begin{equation}\label{ty_solution}
\dot{a}' = \dot{a} \frac{N'}{N}.
\end{equation}
It is somewhat surprising that \eqref{ty_solution} is already the same solution that solve the $ty$-field equation in braneworld Einstein-Hilbert cosmological model \cite{binetruy_1,binetruy_2}. Similarly, variation of \eqref{action_after_assumption} with respect to $b$ give us the $yy$-field equation
\begin{equation}\label{yy_field_eq}
\begin{aligned}
&6\left(kaN + \frac{\ddot{a} a^2}{N} + \frac{\dot{a}^2 a}{N} - \frac{\dot{a}\dot{N}a^2}{N^2} - a'a^2N' - a'^2 a N\right) + \xi_2 a^3 N G_2
\\&+6\xi_4\left[G_4 \left( k a N + \frac{\ddot{a}a^2}{N} + \frac{\dot{a}^2 a}{N} - \frac{\dot{a} \dot{N} a^2}{N^2}\right) + \left(2 G_{4X} \phi'^2 - G_4\right) (a' a^2 N' + a'^2 a N)\right]
\\&- 4\xi_5 G_{5X} \phi'^3 (a'^3 N + 3 a'^2 a N') + \mathcal{C}(t) \phi' =0,
\end{aligned}
\end{equation}
where we have used the definition of $\mathcal{C}(t)$ in \eqref{scalar_field_eq} and set $b=1$. At this point we can revert back to our original metric \eqref{metric}.

Now we will derive the $tt$ and $ij$-field equations from the variation of $N$ and $a$ respectively. By varying the action with respect to $N$ we have
\begin{equation}\label{tt_field_eq}
\begin{aligned}
&6 \left(ka + \frac{\dot{a}^2 a}{N^2}- a'' a^2 - a'^2 a\right)+\xi_2 a^3 G_2 +2\xi_3 a^3 \phi'^2\phi'' \left(3F_{3X} + 2\phi'^2 F_{3XX} \right)  
\\&+6\xi_4\Bigg[\left(2G_{4X} \phi'^2 - G_4 \right) \left(a''a^2 + a'^2a \right) \\&~~~~~+G_4 \left(ka + \frac{\dot{a}^2 a}{N^4} \right) + 2a'a^2\phi' \phi''\left(2G_{4XX}\phi'^2 + G_{4X}\right)\Bigg]
\\&+ \xi_5\Bigg[3 \left(2F_{5X} \phi'^2 + F_5\right) \frac{\phi''}{b}\left(ka + \frac{\dot{a}^2 a}{N^2}\right) \\&~~~- 4 G_{5X} \phi'^3 (a'^3 + 3 a a' a'')  - 6 a'^2 a\phi'^2 \phi''  \left(2G_{5XX}\phi'^2 + 3 G_{5X}\right)\Bigg].\\&= 0.
\end{aligned}
\end{equation}
The variation with respect to $a$ that give us the $ij$-field equation, will be a pretty complicated expression. On the other hand, because our ultimate purpose is not to solve the bulk equations but rather to find the effective Friedmann equation on the brane, it will be clear later that the sole purpose of the $ij$-field equation is to find the appropriate metric junction condition on the brane. Hence in the following, we will only calculate the metric variable $a$ or $N$ terms that is second order in the derivative of $y$. Thus by varying \eqref{action_after_assumption} with respect to $a$, we find
\begin{equation}\label{ij_field_eq}
\begin{aligned}
&- 6 \left(a^2 N'' + 2 a a'' N \right)+6\xi_4\left(2G_{4X} \phi'^2 - G_4\right) \left(2 a'' a N + a^2 N''\right)  \\&-12\xi_5 G_{5X} \phi'^3 \left(a'^2 N' + a' a'' N + a a'' N' + a a' N''\right) + \text{ others} =0,
\end{aligned}
\end{equation}
where "others" refers to any other terms that contain $\phi''$ or first order (or less) derivative of $y$.

Basically, we have obtained all of the bulk field equations. To take the existence of brane into consideration, we need junction conditions on the brane. First, note that the variation of four dimensional brane action $S_b$ with respect to some variable, let say $a$, contributes to the five dimensional bulk field equations a term that looks like the following
\begin{equation}\label{contribution_of_brane}
\frac{\delta S_b}{\delta a} \delta(y).
\end{equation}
A term like this, only comes into play when we evaluate the field equations on the brane, where $y=0$. We thus need to look for terms that contribute Dirac delta function $\delta(y)$ in the bulk field equations.

Now, let $\phi''(y=0) = 0$ so that $\phi'$ is continuous on the brane. Notice that this was the reason that we didn't compute the terms that contain $\phi''$ in \eqref{ij_field_eq}. On the other hand, we only require the continuity of $a$ and $N$ so that neither $a'$ nor $N'$ need to be continuous. Thus $a''$ and $N''$ will contain distributional part $[a'']_{\text{D}}$ and $[N'']_{\text{D}}$ respectively as follows \cite{binetruy_1}
\begin{align}
a'' &= [a'']_{\text{ND}} + [a'']_{\text{D}}, \\
N'' &= [N'']_{\text{ND}} + [N'']_{\text{D}},
\end{align}
while $\text{ND}$ denotes the nondistributional part. For example if $a' = y^2 + |y|$, then $[a'']_{\text{ND}} = 2y$. On the other hand, $[a'']_{\text{D}}$ captures the discontinuity of $a'$ at $y=0$ as follows \cite{binetruy_1}
\begin{equation}
[a'']_{\text{D}} = \Big[a'(y=\epsilon) - a'(y=-\epsilon)\Big]\delta(y),
\end{equation} 
for a small $\epsilon>0$, and similarly for $N$. Finally, by considering the term from brane action \eqref{contribution_of_brane}, we could integrate the $tt$ and $ij$-field equations \eqref{tt_field_eq}, \eqref{ij_field_eq} to obtain the junction conditions for the metric 

\begin{align}
\Bigg[\frac{a'}{a}B_H - \tilde{\alpha} \left(\frac{a'}{a}\right)^2\Bigg]_{y=0} &= - \frac{\kappa_5^2}{6} \rho, \label{cardassian_junction_1}\\
\Bigg[\left(2\frac{a'}{a} + \frac{N'}{N}\right)B_H - \tilde{\alpha} \left[\left(\frac{a'}{a}\right)^2 + 2 \frac{a'}{a}\frac{N'}{N}\right]\Bigg]_{y=0} &=  \frac{\kappa_5^2}{2}  p \label{cardassian_junction_2},
\end{align}
where we have set $\sigma=0$, applied the $\mathbb{Z}_2$ symmetry on the brane 
\begin{equation}
\begin{aligned}
a'(y=\epsilon) &= - a'(y=-\epsilon),\\
N'(y=\epsilon) &= - N'(y=-\epsilon),
\end{aligned}
\end{equation}
and used the following definition
\begin{align}
B_H &\equiv \left[1+\xi_4(G_4-2G_{4X}\phi'^2)\right],\\
\tilde{\alpha} &\equiv -2 \xi_5 G_{5X} \phi'^3.
\end{align}
It can be checked that the previous junction conditions \eqref{cardassian_junction_1}, \eqref{cardassian_junction_2} satisfy the conservation of energy momentum tensor

\begin{equation}\label{conserved_EMT}
\dot{\rho} + 3 \frac{\dot{a}}{a} ( \rho + p) = 0.
\end{equation}

Finally, to derive the junction condition for the scalar field, consider again the scalar field equation, but this time with the contribution from the brane action \eqref{contribution_of_brane}
\begin{equation}\label{junction_scalar_1}
\sqrt{-q}\frac{\partial \mathcal{L}_b}{\partial \phi} \delta(y) -  \frac{d}{dy} \frac{\partial \mathcal{L}}{\partial \phi'}=0. 
\end{equation}
By integrating and applying the $\mathbb{Z}_2$ symmetry to \eqref{junction_scalar_1} we have
\begin{equation}\label{junction_scalar_2}
\left[\frac{\partial \mathcal{L}_b}{\partial \phi}\right]_{y=0} = \left[ \frac{2}{\sqrt{-q}} \frac{\partial \mathcal{L}}{\partial \phi'}\right]_{y=0},
\end{equation}
so that junction condition for scalar field simply say that brane Lagrangian $\mathcal{L}_b$ contains  a term $\ell_b[\phi]$, which is defined as follows
\begin{equation}
\ell_b[\phi] \equiv \phi \left[\frac{2}{\sqrt{-q}} \frac{\partial \mathcal{L}}{\partial \phi'}\right]_{y=0}.
\end{equation}

Now, assume that $\mathcal{L}_5$ Lagrangian is strongly coupled to the rest of the Lagrangian
\begin{equation}\label{cardassian_strong_coupling_cond}
\left|\tilde{\alpha} \frac{a'}{a}\right| \gg B_H,
\end{equation}
so that the solution of the first junction condition for this model \eqref{cardassian_junction_1} can be approximated by 
\begin{equation}\label{cardassian_junction_sol}
\left[\frac{a'}{a}\right]_{y=0} = \left(\frac{\tilde{\beta}}{\tilde{\alpha}}\right)^{1/2}, 
\end{equation}
where
\begin{equation}
\tilde{\beta} \equiv \frac{\kappa_5^2}{6}\rho.
\end{equation}

Now, following Binetruy, et al. \cite{binetruy_2}, using \eqref{ty_solution}, it can be shown that the $yy$ \eqref{yy_field_eq} and $tt$ \eqref{tt_field_eq} field equations can be rewritten into first order differential equations
\begin{equation}
\begin{aligned}
&\dot{\chi} =  \frac{\dot{a}}{3N} \times \frac{\partial \mathcal{L}}{\partial y} = 0,\\
&\chi' = \frac{a'}{3} \times \frac{\partial \mathcal{L}}{\partial N} = 0,
\end{aligned}
\end{equation}
where $\chi$ is defined through the following relation

\begin{equation}\label{cardassian_first_order}
\begin{aligned}
H^2 \left[B_H - \tilde{\alpha} \frac{a'}{a} \right] &= - \frac{k}{a^2} \left[B_H - \tilde{\alpha} \frac{a'}{a} \right] - B_1 \frac{a'}{a} - B_{2} \left(\frac{a'}{a}\right)^2 - B_3 \left(\frac{a'}{a}\right)^3 \\&~~~+ \frac{\chi}{a^4} - B_0,
\end{aligned}
\end{equation}
with the following definitions 
\begin{equation}\label{cardassian_definition_B_i}
\begin{aligned}
B_0 &\equiv \frac{\xi_2}{12}\left(G_2 - 2G_{2X} \phi'^2\right), \\
B_1 &\equiv \frac{2}{3} \xi_3\phi'^3 (3 F_{3X} + 2 \phi'^2 F_{3XX}), \\
B_2 &\equiv -\left[1- \xi_4 \left(4 G_{4X} \phi'^2 + 2 G_{4XX} \phi'^4 - G_4\right)\right], \\
B_3 &\equiv -\frac{2}{3} \xi_5 \phi'^3 ( 2G_{5XX} \phi'^2 + 5 G_{5X}).
\end{aligned}
\end{equation}

Finally, from \eqref{cardassian_first_order} and \eqref{cardassian_junction_sol}, we secure the Friedmann equation for the strongly coupled $\mathcal{L}_5$
\begin{equation}\label{cardassian_friedmann_eq}
H^2= - \frac{k}{a^2} + \frac{B_1}{\tilde{\alpha}} + \left(\frac{\kappa_5^2}{6\tilde{\alpha}^3}\right)^{1/2} B_2 \rho^{1/2} + \frac{\kappa_5^2}{6 \tilde{\alpha}^2} B_3 \rho -  \left(\frac{6}{\tilde{\alpha} \kappa_5^2}\right)^{1/2} \left(\frac{\chi}{a^4}-B_0\right) \rho^{-1/2}.
\end{equation}

From \eqref{cardassian_friedmann_eq}, we have shown that Horndeski Lagrangian  \eqref{action_after_assumption} with strongly coupled $\mathcal{L}_5$ \eqref{cardassian_strong_coupling_cond} is one of the specific bulk energy momentum tensor $T_{AB}$ in braneworld scenario \cite{chung1999cosmological} that will generate Cardassian terms \cite{freese2002cardassian} $\rho^n$ with $n=\pm 1/2$ in its four dimensional effective Friedmann equations. Furthermore, the latest combined observational evidence in 2017 from BAO, CMB, SNIa, $f_{\sigma 8}$, and $H_0$ value observation, has given the following constraints for the polytropic Cardassian \eqref{cardassian_ansatz_friedmann_poly} \cite{zhai2017evaluation}
\begin{equation}\label{cardassian_batas_parameter_m_n}
m = 1.1^{+0.8}_{-0.4}, \hspace{5mm} n=0.02^{+0.25}_{-0.41},
\end{equation}
so that the $n=-1/2$ term lies quite close to the observed value. However, it should be noted that the modified Friedmann equation \eqref{cardassian_friedmann_eq} is more general than \eqref{cardassian_ansatz_friedmann_poly} and thus requires a numerical evaluation on its own.

\section{Modified Friedmann equations for the weak $\mathcal{L}_5$ coupling}
In the previous section, we saw that the strongly coupled $\mathcal{L}_5$ case produced Cardassian terms which is basically a lower energy correction for the matter term. In this section, we will see how the weakly coupled $\mathcal{L}_5$ case generates among others the cubic matter term $\rho^3$ which is a high energy correction. 

We start by assuming that $\mathcal{L}_5$ Lagrangian is weakly coupled to the rest of the Lagrangian so that $\xi_5$ is small and thus $\xi_5^2$ can be abandoned.  Next, from \eqref{Horndeski_Lagrangian_geometric}, it can be seen that the $\mathcal{L}_4$ has an almost identical expression with $R$ \eqref{Ricci_scalar_projection}, differing only with some scalar field function coefficients. Thus, in order to make our result tidier, we will also abandon $\mathcal{L}_4$. In the rest of the section, we will identify the correction terms provided by the Friedmann equation obtained from general scalar-tensor braneworld model \eqref{action_after_assumption} where the $\xi_4=0$ and  $\xi_5$ is small.

Firstly, the junction conditions for $\xi_4=0$ are
\begin{align}
\Bigg[\frac{a'}{a} + \alpha \left(\frac{a'}{a}\right)^2\Bigg]_{y=0} &= - \frac{\kappa_5^2}{6} \rho \label{tt_junction},\\
\Bigg[2\frac{a'}{a} + \frac{N'}{N} + \alpha \left[\left(\frac{a'}{a}\right)^2 + 2 \frac{a'}{a}\frac{N'}{N}\right]\Bigg]_{y=0} &=  \frac{\kappa_5^2}{2}  p, \label{ij_junction}
\end{align}
where
\begin{equation}
\alpha \equiv 2\xi_5 \phi'^3 G_{5X}.
\end{equation}
It can be checked that these junction conditions satisfy the conservation of energy momentum tensor \eqref{conserved_EMT}.

Now, assume that $\mathcal{L}_5$ Lagrangian is weakly coupled to the rest of the Lagrangian, so that $\xi_5$ is small. From \eqref{tt_junction}, expansion of $a'/a$ to various order of $\alpha$ give us the explicit solution for the junction conditions 
\begin{equation}\label{junction_solution}
\left[\frac{a'}{a}\right]_{y=0} = - \frac{\kappa_5^2}{6}\rho \left(1+ \alpha\frac{\kappa_5^2}{6}\rho\right).
\end{equation}

Next, analogous to \eqref{cardassian_first_order} for $\xi_4=0$ we have the following definition for $\chi$
\begin{equation}\label{first_order}
\begin{aligned}
\chi &= \left[ka^2 -(a'a)^2 + \frac{(\dot{a}a)^2}{N^2}\right]+ \frac{\xi_2}{12}a^4(G_2 - 2G_{2X}\phi'^2) +\frac{2}{3} \xi_3 \phi'^3 a' a^3 (3F_{1X} + 2 \phi'^2 F_{1XX}) \\&~~+ 2 \xi_5 \phi'^3 G_{5X} (kaa' + a a' \dot{a}^2) - \frac{2}{3} \xi_5 \phi'^3 a a'^3 (2 G_{5XX} \phi'^2 + G _{5X}).
\end{aligned}
\end{equation}
By evaluating \eqref{first_order} on the brane using junction conditions solution \eqref{junction_solution}, we get the Friedmann equation for this model up to linear order of $\alpha$
\begin{equation}\label{friedmann_horndeski}
\begin{aligned}
H^2 = - \frac{k}{a^2} + \frac{\kappa_5^2}{6} A_1 \rho + \frac{\kappa_5^4}{36} A_2 \rho^2 + \frac{\kappa_5^6}{216} A_3 \rho^3 + \frac{\chi}{a^4}\left(1+ \frac{\alpha\kappa_5^2}{6}\rho\right) + A_0,
\end{aligned}
\end{equation}
where $N(y=0)=1$ has been taken, and the following definitions have been used
\begin{equation}\label{parameter_Ai}
\begin{aligned}
A_0 &\equiv - \frac{\xi_2}{12}(G_2 - 2G_{2X} \phi'^2),\\
A_1 &\equiv \frac{2}{3} \xi_3 \phi'^3(3F_{3X}+2\phi'^2F_{3XX})  -  \frac{\alpha\xi_2}{12}(G_2 - 2G_{2X}\phi'^2),\\
A_2 &\equiv 1+\frac{4\alpha}{3}  \xi_3 \phi'^3(3F_{3X}+2\phi'^2F_{3XX}), \\
A_3 &\equiv \frac{4}{3}\left(\alpha - \xi_5 \phi'^5 G_{5XX} \right).
\end{aligned}
\end{equation}
For consistency, it can checked that by taking  
\begin{equation}
\xi_2 = - 2 \kappa_5^2, \hspace{3mm} \xi_3, ~\xi_5 = 0, \hspace{3mm} G_2 = \Lambda_5,
\end{equation}
the Friedmann equation in this model \eqref{friedmann_horndeski} go back to the braneworld Einstein-Hilbert model \eqref{friedmann_1_braneworld_EH} with $\sigma=0$
\begin{equation}
H^2 = -\frac{k}{a^2} + \frac{\kappa_5^4}{36}\rho^2 + \frac{\chi}{a^4} + \frac{\kappa_5^2}{6}\Lambda_5.
\end{equation} 

This model Friedmann equation \eqref{friedmann_horndeski} contain some new correction terms which are not present in the Einstein-Hilbert braneworld model \eqref{friedmann_1_braneworld_EH}. Firstly, notice that every terms coefficient is some sort of function build from the scalar field. In fact, we can identify the four dimensional Einstein's kappa constant as follows
\begin{equation}\label{kappa_4d_identification}
\frac{\kappa_4^2}{3} = \frac{\kappa_5^2}{6}A_1.
\end{equation}
From \eqref{kappa_4d_identification}, it can be inferred that $A_1^{-1}$ is proportional to radius of the extra dimension. Next, we can see that in addition to quadratic matter term, we also have another high energy correction term, the cubic matter. Interestingly, we also have a term proportional to $\chi a^{-4} \rho$ which mediate the interaction between normal matter and the dark radiation. This interaction though is small because $\alpha$ is small from the assumption that $\mathcal{L}_5$ is weakly coupled.  Lastly, as usual, we still have the cosmological constant term $A_0$ coming from the the scalar field $\phi$ evaluated on the brane. Therefore, assuming $\xi_2<0$, in the cosmological constant domination phase, the universe in this model is undergoing de Sitter expansion
\begin{equation}
a(t) = a_0 \exp\left({\sqrt{A_0}t}\right).
\end{equation}
Also notice that for $\xi_2 >0$, the scale factor is oscillating: $a(t) \propto \exp\left(i \sqrt{-A_0} t\right)$.

Before finding the constraint for the value of $A_1$, $A_2$, and $A_3$, we will transform the Friedmann equation \eqref{friedmann_horndeski} into an explicit form in term of Hubble function. Define the following density parameters 
\begin{equation}\label{density_param}
\begin{aligned}
\Omega_{\rho,0} = \frac{\kappa_4^2}{3} \frac{\rho_0}{H_0^2}, \hspace{3mm} \Omega_{\chi,0} =  \frac{\chi}{H_0^2}, \hspace{3mm} \Omega_{k,0} = -\frac{k}{H_0^2}, \hspace{3mm} \Omega_{\Lambda,0} = \frac{A_0}{H_0^2}.
\end{aligned}
\end{equation}
In the previous definitions \eqref{density_param}, $\rho_0$ is the sum of radiation and dust, evaluated in the present epoch
\begin{equation}
\Omega_{\rho,0} = \Omega_{r,0} + \Omega_{m,0}, \hspace{3mm} \text{where} \hspace{3mm} \Omega_{r,0} = \frac{\kappa_4^2}{3} \frac{\rho_{r,0}}{H_0^2}, \hspace{3mm} \Omega_{m,0} = \frac{\kappa_4^2}{3} \frac{\rho_{m,0}}{H_0^2}.
\end{equation}

Now, assuming that the spatial curvature of the universe is $\Omega_{k,0} = 0$, in the epoch of single matter domination with equation of state $p=w\rho$, the Friedmann equation \eqref{friedmann_horndeski} can be recast into
\begin{equation}\label{friedmann_horndeski_in_z}
\begin{aligned}
H(z) &= H_0 \Big[\Omega_{\rho,0} (1+z)^{3(w+1)} + H_0^2 A_1^{-2} A_2 \Omega_{\rho,0}^2 (1+z)^{6(w+1)} \\&~~+ H_0^4 A_1^{-3} A_3\Omega_{\rho,0}^3 (1+z)^{9(w+1)} + \Omega_{\chi,0}(1+z)^4 \\&~~+ \alpha H_0^2 A_1^{-1} \Omega_{\chi,0}\Omega_{\rho,0} (1+z)^{(3w+7)} + \Omega_{\Lambda,0}\Big]^{1/2},
\end{aligned}
\end{equation}
with $z=a^{-1}-1$ is the redshift factor.

Now we are ready to constraint $A_1$, $A_2$, and $A_3$ using  big bang nucleosynthesis (BBN) constraint. BBN constraint say that  the contribution of high energy correction term \eqref{friedmann_horndeski_in_z} must be negligible relative to the contribution of linear matter contribution before the BBN, where $z_{\text{BBN}} \simeq 4 \times 10^8$ \cite{maartens_2010}. Because when $z=z_{\text{BBN}}$ the universe is dominated by radiation, $\Omega_{\rho,0} \approx \Omega_{r,0}$, then BBN constraint give the following conditions
\begin{equation}
\begin{aligned}
\theta_1 &\equiv \frac{H_0^2 A_1^{-2} A_2 \Omega_{r,0}^2 (1+z_{\text{BBN}})^{8}}{\Omega_{r,0} (1+z_{\text{BBN}})^{4}} = H_0^2 A_1^{-2} A_2 \Omega_{r,0} (1+z_{\text{BBN}})^{4} \ll 1,\\
\theta_2 & \equiv \frac{H_0^4 A_1^{-3} A_3\Omega_{r,0}^3 (1+z_{\text{BBN}})^{12}}{\Omega_{r,0} (1+z_{\text{BBN}})^{4}} = H_0^4 A_1^{-3} A_3 \Omega_{r,0}^2 (1+z_{\text{BBN}})^{8} \ll 1. 
\end{aligned}
\end{equation}
Note that both $\theta_1$ and $\theta_2$ are dimensionless. Now by taking $H_0 = 67$ km s${}^{-1}$ Mpc${}^{-1}$ $\approx 2\cdot 10^{-18}$ s${}^{-1}$ (as in Ref. \cite{planck_2015}) and $\Omega_{r,0} \approx 5 \cdot 10^{-5}$, in SI unit
\begin{equation}
\begin{aligned}
&\theta_1 \ll 1 \implies A_1^{-2} A_2 \ll 10^5, \\
&\theta_2 \ll 2 \implies A_1^{-3} A_3 \ll 10^{11}.
\end{aligned}
\end{equation}  

From definition \eqref{parameter_Ai} and small $\alpha$ condition, then $A_2 \approx 1$ s${}^{2}$ m${}^{-2}$. Thus from the condition of small $\theta_1$, we have $A_1 \gg 10^{-2.5}$ m${}^{-1}$. Now, because from \eqref{kappa_4d_identification} $A_1^{-1}$ is proportional to extra dimension radius, it can be inferred that the condition for $\theta_1$ simply says that the extra dimension radius must be smaller than 10${}^{2.5}$ m, which is a natural condition. 

Similarly, condition for $\theta_2$ gives the upper bound for $A_3$, that is $A_3 \ll 10^{11} A_1^{3}$. In principle, the radius of extra dimension can be very small, even as small as $M_\text{P}^{-1}$ as in Randall-Sundrum I model \cite{randall_1}. Therefore $A_1$ can have an arbitrarily high value. For example, if we take $A_1 \approx 10^{-2}$ m${}^{-1}$, condition for $\theta_2$ gives us the condition that $A_3 \ll 10^{2}$ s${}^{2}$ m${}^{-2}$. From the definition of $A_3$ \eqref{parameter_Ai}, the previous condition is a natural one, because in this model we assume that $\alpha$ is small. In conclusion, this model provides the high energy correction term and supports the BBN process.

This characteristic is not so common, and in fact, is not shared by the braneworld Einstein-Hilbert model \eqref{friedmann_1_braneworld_EH}. The Friedmann equation in that model for $k=0$ can be recast into
\begin{equation}
H = H_0 \left[\Omega_{\rho,0}(1+z)^{3(1+w)}+ \frac{3H_0^2}{2 \kappa_4^2 \sigma}\Omega_{\rho,0}^2 (1+z)^{6(1+w)}+ \Omega_{\chi,0}(1+z)^4 + \Omega_{\Lambda,0}\right]^{1/2},
\end{equation}
with the following identification
\begin{equation}
\begin{aligned}
\frac{\kappa_4^2}{3}&=\frac{\kappa_5^4\sigma}{18}, \\
\Omega_{\Lambda,0} &= \frac{\kappa_5^2}{6}\left(\frac{\kappa_5^2 \sigma^2}{6}+ \Lambda_5\right).
\end{aligned}
\end{equation}
BBN condition for this model requires an unnaturally high value of brane tension
\begin{equation}
\sigma \gg \frac{3H_0^2}{2 \kappa_4^2} \Omega_{r,0} (1+z_{\text{BBN}})^{4} \approx 10^{20} ~\frac{\text{kg}}{\text{m s}{}^{2}}.
\end{equation}

To close this section, we will give a visualization of how the dark radiation might affect the evolution of the universe. In principle, dark radiation is just an ordinary radiation, but this time with no requirement for the value to be nonnegative. This in fact, can make the total sum of energy density parameter of the other component of the universe to be greater than one even in a flat universe. To be more precise, consider the following models

\begin{equation}\label{three_model}
\begin{aligned}
H_\text{EH}(z) &= H_0 \left[\Omega_{m,0}(1+z)^{3}+ \Omega_{\Lambda,0}\right]^{1/2},\\
H_\text{BW}(z) &= H_0 \left[\Omega_{m,0}(1+z)^{3}+ \frac{3H_0^2}{2 \kappa_4^2 \sigma}\Omega_{m,0}^2 (1+z)^{6}+ \Omega_{\chi,0}(1+z)^4 + \Omega_{\Lambda,0}\right]^{1/2},\\
H_\text{HD}(z) &= H_0 \Big[\Omega_{m,0} (1+z)^{3} + H_0^2 A_1^{-2} A_2 \Omega_{m,0}^2 (1+z)^{6} + H_0^4 A_1^{-3} A_3\Omega_{m,0}^3 (1+z)^{9} \\&\hspace{12mm}+ \Omega_{\chi,0}(1+z)^4 + \alpha H_0^2 A_1^{-1} \Omega_{\chi,0}\Omega_{m,0} (1+z)^{7} + \Omega_{\Lambda,0}\Big]^{1/2}, \\
\end{aligned}
\end{equation}
where $H_\text{EH}(z)$, $H_\text{BW}(z)$, and $H_{\text{HD}}(z)$ refer to conventional four dimensional model, braneworld model with Einstein-Hilbert action \eqref{friedmann_1_braneworld_EH}, and general scalar-tensor braneworld model \eqref{friedmann_horndeski}. In those expressions, we also have assumed that $\Omega_{k,0} = 0$ and $\Omega_{r,0} = 0$

\begin{figure}[h!]
\centering
\includegraphics[scale=0.8]{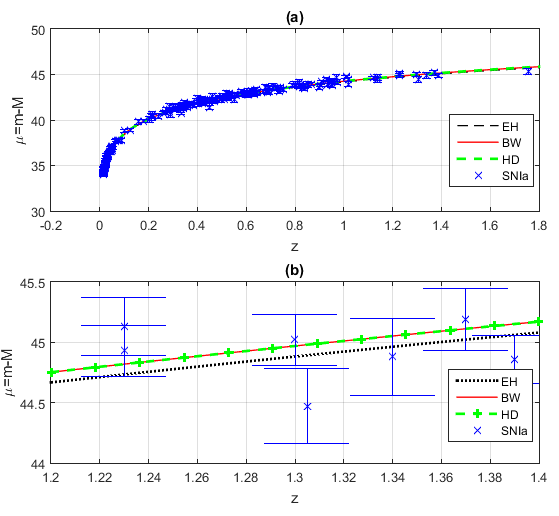}
\caption[Comparison of Hubble diagram for conventional four dimensional model, braneworld Einstein-Hilbert model, and general scalar-tensor braneworld model with SNIa data Davis, et al.(2007)]{\small Comparison of Hubble diagram for conventional four dimensional model (EH), braneworld Einstein-Hilbert model (BW), and general scalar-tensor braneworld model (HD) with SNIa data, Davis, et al.(2007) \cite{davis_scrutinizing,vassey_observational,riess_new_hubble}. For  EH, we use the cosmological parameter from Planck 2015 \cite{planck_2015} $H_0 = 67$ km s${}^{-1}$ Mpc${}^{-1}$, $\Omega_{m,0} = 0.308$, $\Omega_{\Lambda,0} = 0.692$. For BW and HD, we use $H_0=66.157$ km s${}^{-1}$ Mpc${}^{-1}$, $\Omega_{m,0} = 0.3453$, $\Omega_{\Lambda,0} = 0.6568$ and $\Omega_{\mathcal{E},0} = -0.03$. The numerical value for the rest are  $A_1, A_3, \alpha = 10^{-3}$, $A_2 = 1$, and $\sigma = 10^{22}$. The chi-square value $\chi_\epsilon^2$ for EH, BW and HD are 207.8483, 197.1633, and 197.1633 respectively. By taking $\mathcal{N} = 192 -4 = 188$, the value of $\chi_\epsilon^2 \mathcal{N}^{-1}$ for EH, BW and HD are  1.1056, 1.0487, 1.0487 respectively.}
\label{fig:hubble_diag}
\end{figure}

The previous models will be compared with $N_{\text{obs}} = 192$ SNIa data compiled by Davis, et al. in 2007 \cite{davis_scrutinizing,vassey_observational,riess_new_hubble}. Observation data are given in terms of modulus distance ($\mu$) against redshift ($z$). From the various expressions for $H(z)$ in \eqref{three_model}, the modulus distance can be calculated as follows
\begin{equation}
\mu(z) = 5 \log \left[\frac{(1+z)c}{\text{10 pc}}\int_0^z \frac{dz'}{H(z')}\right],
\end{equation}
where $c$ is the speed of light in vacuum and pc is parsec. A good model is one with $\chi_\epsilon^2 \mathcal{N}^{-1} \approx 1$, where the chi-squared error $\chi_\epsilon^2$ is defined as follows \cite{numericalrecipes_1992} 
\begin{equation}\label{{epsilon_square}}
\chi_\epsilon^2 = \sum_{i=1}^{N_{\text{obs}}} \frac{[\mu(z_i) - \mu_{\text{obs}}(z_i)]^2}{\sigma_i^2},
\end{equation}
where $\mu_{\text{obs}}(z_i)$ is the modulus distance obtained from observation for redshift factor $z=z_i$ and error $\sigma_i$, while $\mathcal{N}$ is the difference between number of data $N_{\text{obs}}$ with the free parameter of the model. The difference between evolution of the three models \eqref{three_model} and the observational data is given in figure \ref{fig:hubble_diag}.

It should be noted that supernova is a phenomenon with a low $z$. For example, the data used in this article has $z_{\text{max}} = 1.8$.
Thus the high energy correction terms provided by modified Friedmann equation can't be detected via supernova observation. It can also be inferred that, practically all the three models only have three free parameters $H_0$, $\Omega_{m,0}$, and $\Omega_{\Lambda,0}$, while $\Omega_{\mathcal{E},0}$ can be approximated from $\Omega_{\mathcal{E},0} \approx 1 - \Omega_{m,0} - \Omega_{\Lambda,0}$. From the Hubble diagram of figure \ref{fig:hubble_diag}, it can be seen that  braneworld Einstein-Hilbert \eqref{friedmann_1_braneworld_EH} and general scalar-tensor braneworld model \eqref{friedmann_horndeski} that has different high order correction, can't be distinguished. Meanwhile,  the four dimensional conventional model can be distinguished from the others by the contribution of dark radiation term $\Omega_{\mathcal{\chi},0}$.

\section{Conclusion and outlook}
A braneworld cosmological model in general scalar-tensor action comprises of various Horndeski Lagrangian have been investigated. The derivation of the corresponding field equations has been given. The resulting Friedmann equation in this model, in the case of strongly coupled $\mathcal{L}_5$ model produces the Cardassian term $\rho^n$ with $n=\pm 1/2$, which can served as alternative explanation for the accelerated expansion phase of the universe. The latest combined observational facts from BAO, CMB, SNIa, $f_{\sigma_8}$, and $H_0$ value observation, suggest that the $n=-1/2$ term lies quite close to the constrained value. On the other hand, the weakly coupled $\mathcal{L}_5$ case has several new correction terms, e.g. the cubic term $\rho^3$ and the dark radiation-matter interaction term $\propto \chi a^{-4} \rho$. Furthermore, this model provides a cosmological constant constructed from the bulk scalar field, requires no brane tension, and supports the BBN  constraint naturally. For the future works, on the computational side it is interesting to consider a parameter numerical fitting for the specific Cardassian model provided in this paper \eqref{cardassian_friedmann_eq}. On the other hand, one might be interested to consider braneworld cosmological model with a more general action than Horndeski, for example the beyond Horndeski model or Proca theory. 

\section*{Acknowledgments}
AS and FPZ gratefully acknowledge  the  support
from Ministry of Research, Technology, and Higher Education
of the Republic of Indonesia for the PDUPT Research Grant.


\end{document}